\documentclass{ws-ijmpcs}

\begin{document}

\markboth{Eug\^enio R. Bezerra de Mello}
{Wightman functions}

%
\catchline{}{}{}{}{}
%

\title{\bf WIGHTMAN FUNCTIONS IN DE SITTER AND ANTI-DE SITTER SPACETIMES IN THE PRESENCE OF COSMIC STRINGS}

\author{EUG\^ENIO R. BEZERRA DE MELLO\footnote{emello@fisica.ufpb.br}}

\address{Departamento de F\'{\i}sica-CCEN, Universidade Federal da Para\'{\i}ba\\
J. Pessoa, PB, 58.059-970, Brazil}

\maketitle

\begin{history}
\received{Day Month Year}
\revised{Day Month Year}
\end{history}

\begin{abstract}
In this paper we evaluate the Wightman functions associated with a massive quantum scalar field in de Sitter and anti-de Sitter spacetimes in the presence of a cosmic string. Having these functions we calculate the corresponding renormalized vacuum expectation values of the field squared and present the behavior of the contributions induced by the cosmic string as function of the proper distance to it for different values of the parameter which codify the presence of this linear topological defect.
\end{abstract}

\ccode{PACS numbers: 98.80.Cq, 11.10.Gh, 11.27.+d}

\section{Introduction}	
Phase transitions in the early Universe have several cosmological consequences and provide an important link between particle physics and cosmology. In particular, different types of topological defects may have been created by the vacuum phase transitions after Planck time.\cite {Kibble,V-S} Among them the cosmic strings are of special interest. The gravitational field produced by a cosmic string may be approximated by a
planar angle deficit in the two-dimensional sub-space. In quantum field theory the corresponding non-trivial topology of this sub-space induces non-zero vacuum expectation values for physical observables. In this context, the vacuum expectation values (VEVs) of the energy-momentum tensor have been calculated for scalar\cite{scalar}\cdash\cite{scalar4} and fermionic\cite{ferm}\cdash\cite{ferm3} fields.

De Sitter (dS) and anti-de Sitter (AdS) spaces are curved spacetimes which have been most studied in quantum field theory during the past two decades. The main reason resides in the fact that they are curved maximally symmetric spacetimes and several physical problem can be exactly solvable on these backgrounds. Specifically the importance of dS spacetime has increased by the appearance of the inflationary cosmology scenario.\cite{Linde} In great number of inflationary models, approximated dS spacetime is employed to solve relevant problems in standard cosmology. As to the AdS space, one of its more important applications arises in braneworld scenario. The large extra dimension provides a solution to the hierarchy problem between the gravitational and electroweak mass scales.\footnote{See Ref.~\refcite{Brax03} for reviews on braneworld gravity and cosmology} The vacuum polarizations effects on the background of dS and AdS spacetimes, have been discussed, respectively, in\cite{Cher68}\cdash\cite{Dolg06} and in\cite{Burg85}\cdash\cite{Cald99} for fields of different spins.

In this paper, mainly supported by two previous publications,\cite{Mello1,Mello2} we shall analyze the vacuum polarization effects associated with a massive scalar quantum field in the dS and AdS spacetimes in the presence of a cosmic string. The presence of the string is the new ingredient in the analysis. Because the pure dS or AdS spacetimes are maximally symmetric spaces, the vacuum polarizations effects associated with scalar quantum fields do not depend on the spacetime point. On the other hand, considering the presence of a cosmic string, the corresponding quantities present explicitly dependences on the proper distance to the string. This is basically the content of this paper. To investigate the influence of a cosmic string on the VEVs of the field squared. In order to do that one important quantity to be used is the positive frequency Wightman function, which by its turn depends on the knowledge of the complete set of normalized solution of the corresponding Klein-Gordon equations.

\section{Klein-Gordon equation}

The main objective of this section is to obtain a complete set of orthonomarlized solutions of the Klein-Gordon equation for a massive field in a four-dimensional dS and AdS spacetimes in presence of a cosmic string. These functions are important in the calculation of the corresponding Wightman functions. 

We shall develop the calculations considering separately the two spacetimes. We start with the dS case.

\subsection{dS spacetime in the presence of a cosmic string}\label{dS}

The geometry associated with the cosmic string in dS spacetime can be given by using cylindrical coordinates through the following line element:
\begin{equation}
ds^{2}=-dt^{2}+e^{2t/\alpha }(dr^{2}+r^{2}d\phi^{2}+dz^{2}) \ ,  \label{ds1-dS}
\end{equation}
where $r\geq 0$ and $\phi \in \lbrack 0,\ 2\pi /q]$ define the coordinates on the conical two-geometry, $(t,\ z)\in (-\infty ,\ \infty )$, and the parameter $\alpha$ is related with with the cosmological constant and Ricci scalar by the formulas
\begin{equation}
\Lambda =\frac3{\alpha ^{2}} \ ,\ R=\frac{12}{\alpha ^{2}} \ .
\end{equation}
The parameter $q$ bigger than unity, codifies the presence of the cosmic string. 

For further analysis, in addition to the synchronous time coordinate $t$, it is more convenient to introduce the conformal time $\tau $\ according to
\begin{equation}
\tau =-\alpha e^{-t/\alpha}\ ,\ -\infty <\ \tau \ <\ 0\ .
\end{equation}
In terms of this coordinate the above line element takes the form
\begin{equation}
ds^{2}=(\alpha /\tau)^{2}(-d\tau^{2}+dr^{2}+r^{2}d\phi^{2}+dz^{2}) \ .  \label{ds2-dS}
\end{equation}

The field equation that will be considered is
\begin{equation}
(\nabla_\mu\nabla^\mu-m^{2}-\xi R)\Phi (x)=0 \ ,  \label{KG}
\end{equation}
where $\xi $ is an arbitrary curvature coupling constant. The solution of the Klein-Gordon equation in the coordinate system defined by (\ref{ds2-dS}) is:\cite{Mello1}
\begin{equation}
\Phi _{\sigma}(x)=C_{\sigma }\eta ^{3/2}H_{\nu }^{(1)}(\lambda\eta)J_{q|n|}(pr)e^{ikz+in\phi }\ ,  \label{sol1-dS}
\end{equation}
with $\lambda =\sqrt{p^{2}+k^{2}}$, $\eta=-\tau$ and
\begin{equation}
\ \nu =\sqrt{9/4-12\xi-m^{2}\alpha ^{2}} ,\ \ n=0,\pm 1,\pm 2,\ ...
\end{equation}
In (\ref{sol1-dS}), $H_{\nu}^{(1)}$ and $J_{\nu}$ represent the Hankel and Bessel functions respectively, and $\sigma \equiv (p,\ {k},\ n)$ the set of quantum numbers, being $p\in \lbrack 0,\ \infty )$. The coefficient $C_{\sigma }$ can be found by the orthonormalization condition
\begin{equation}\label{norm}
i\int d^3x\sqrt{|g|}g^{00}[\Phi _{\sigma}(x)\partial_{t}\Phi _{\sigma^{\prime}}^{\ast}(x)-\Phi _{\sigma ^{\prime}}^{\ast}(x)\partial_{t}\Phi_{\sigma}^{\ast}(x)]=\delta_{\sigma ,\sigma ^{\prime}}\ ,
\end{equation}
where the integral is evaluated over the spatial hypersurface $\eta =$const, and $\delta _{\sigma ,\sigma^{\prime }}$ represents the Kronecker-delta for discrete index and Dirac-delta function for continuous ones. This leads to
\begin{equation}
C_\sigma=\frac1{4\alpha}\sqrt{\frac{qp}\pi}\ .  \label{coef-dS}
\end{equation}
In this analysis we shall use $\nu$ as being a real number.

\subsection{AdS spacetime in the presence of a cosmic string}\label{AdS}

In cylindrical coordinates, the geometry associated with a cosmic string in a four-dimensional AdS spacetime is given by the line element below (considering a static string along the $y$-axis):
\begin{equation}
ds^{2}=e^{-2y/a}(-dt^{2}+dr^{2}+r^{2}d\phi ^{2})+dy^{2}\ ,  \label{ds1-AdS}
\end{equation}%
where $r\geqslant 0$ and $\phi \in \lbrack 0,\ 2\pi /q]$ define the coordinates on the conical geometry, $(t,\ y)\in (-\infty ,\ \infty )$. Note that the curvature scale $a$ is related to the cosmological constant, $\Lambda $, and the Ricci scalar, $R$, by the formulas
\begin{equation}
\Lambda =-\frac3{a^{2}},\ \ R=-\frac{12}{a^{2}}\ .
\label{LamR}
\end{equation}
By using \textit{Poincar\'{e}} coordinate defined by $z=ae^{y/a}$, the line element above is written in the form conformally related to the line element associated with a cosmic string in Minkowski spacetime:
\begin{equation}
ds^{2}=(a/z)^{2}(-dt^{2}+dr^{2}+r^{2}d\phi ^{2}+dz^{2})\ .  \label{ds2-AdS}
\end{equation}
For the new coordinate one has $z\in \lbrack 0,\ \infty )$. Limiting values $z=0$ and $z=\infty$ correspond to the AdS boundary and horizon, respectively.

Here also, the field equation is given by \eqref{KG}. In the coordinate system defined by (\ref{ds2-AdS}), the complete set of normalized solutions, according to \eqref{norm}, and regular at the boundary $z=0$, is characterized by the set of quantum number $\sigma =(\lambda ,\ n,\ p \ ,{k})$ and is given by:
\begin{equation}
\Phi_{\sigma}(x)=\sqrt{\frac{q p\lambda}{4\pi a^2E}}\ z^{3/2}J_{\nu}(\lambda z)J_{|n|q}(pr)e^{i(nq\phi +kz-Et)}\ ,  \label{sol1-AdS}
\end{equation}
with
\begin{equation}
(\lambda ,\ p)\in \lbrack 0,\infty )\ ,\ k \in (-\infty ,\ \infty ),\ n=0,\pm 1,\pm 2,\ ...  \label{Ranges}
\end{equation}
In (\ref{sol1-AdS}), $J_{\mu }(u)$ represents the Bessel function, $E=\sqrt{\lambda ^{2}+p^{2}+{k}^{2}}$ and
\begin{equation}
\nu =\sqrt{9/4-12\xi+m^{2}a^{2}} \ .  \label{nu}
\end{equation}

\section{Wightman function}

The main objective of this section is to obtain the positive frequency Wightman functions associated with a massive scalar field in a four-dimensional dS and AdS spacetimes in presence of a cosmic string. These functions are important in the calculations of vacuum polarization effects.

We shall employ the mode-sum formula to calculate the positive frequency
Wightman functions as shown below:
\begin{equation}
G(x,x^{\prime })=\sum_{\sigma}\Phi_{\sigma}(x)\Phi_{\sigma}^{\ast}(x^{\prime})\ .  \label{Green}
\end{equation}

\subsection{dS case}
Substituting \eqref{sol1-dS} and \eqref{coef-dS} into the mode sum \eqref{Green}, in Ref.~\refcite{Mello1} we have shown that the Wightman function can be expressed as:
\begin{equation}
G(x,x^{\prime})=\frac{q\ (\eta \eta ^{\prime })^{3/2}}{\pi^{5/2}\alpha ^{2}}\ \sideset{}{'}{\sum}_{n=0}^{\infty }\cos (nq\Delta\phi )\int_{0}^{\infty }du\,u^{1/2}e^{-\gamma u}I_{qn}(2rr^{\prime}u) K_{\nu }(2\eta \eta ^{\prime }u) \ ,  \label{WF-dS}
\end{equation}
with
\begin{eqnarray}
\gamma=\Delta z^2+r^2+r'^2-\eta^2-\eta'^2 \ .
\end{eqnarray}
In \eqref{WF-dS}, $I_\mu$ and $K_\mu$ are modified Bessel functions, and the prime means that the term $n=0$ should be halved.

The expression for the scalar Wightman function, Eq. (\ref{WF-dS}), becomes a simpler one when $q$ is an integer number. 
Although this case may not be realistic, the results found in this situation may shed light on the behavior of the vacuum polarizations for a general value of $q$. In this case by using the formula\cite{Pru,Spinelly}
\begin{equation}
\sideset{}{'}{\sum}_{n=0}^{\infty }\cos (nq\Delta \phi )I_{qn}(z)= \frac{1}{2q}\sum_{k=0}^{q-1}\exp [z\cos(\Delta\phi+2\pi
k/q)] \ ,  \label{Sumn1}
\end{equation}
the integral over $u$ is evaluated with the help of Ref.~\refcite{Pru}:
\begin{eqnarray}
\int_{0}^{\infty }du\,u^{1/2}e^{-\Delta x_k^2u}K_{\nu }(2\eta \eta ^{\prime }u)&=&\sqrt{\frac{\pi }{4\eta \eta ^{\prime }}}\frac{\Gamma \left( \frac{3}{2}-\nu\right) \Gamma \left( \frac{3}{2}+\nu \right)}{[(2\eta \eta ^{\prime})^{2}-\Delta x_k^{4}]^{1/2}}\nonumber\\
&&\times P_{\nu -1/2}^{-1}(\Delta x_k^2/2\eta \eta ^{\prime }) \ ,
\label{intform2} 
\end{eqnarray}
where $P_{\mu}^{\nu}$ represents the associated Legendre function of the first kind and
\begin{eqnarray}
\Delta x_k^2=\Delta{z}+r^{2}+r^{\prime 2}-2rr\cos (\Delta \phi +2\pi k/q)-(\Delta \eta )^{2}-2\eta \eta ^{\prime} \ .  \label{c}
\end{eqnarray}

So the Wightman function can be expressed in terms of $q$ images of the Wightman functions in dS: 
\begin{align}
\label{Greenk-dS}
G(x',x)=\sum_{k=0}^{q-1}G_k(x',x) \ ,
\end{align}
with
\begin{align}
\label{gk}	G_k(x',x)=\frac{\Gamma\left(\frac32+\nu\right)\Gamma\left(\frac32-\nu\right)}{8\alpha^2\pi^2(1-u_k^2)^{1/2}} P_{\nu-1/2}^{-1}(u_k) \ ,
\end{align}
being
\begin{equation}
u_{k}=\frac{\Delta x_k^2}{2\eta\eta'}=-1+\frac{\Delta{z}+r^{2}+r'^2-2rr'\cos (\Delta\phi +2\pi k/q)-(\Delta \eta )^{2}}{2\eta \eta ^{\prime }}.
\end{equation}

The $k=0$ component of (\ref{Greenk-dS}) is divergent at the coincidence limit, $x'\to x$ ($u_0\to{-1}$), and represents the Wightman function in a pure dS background.

The analysis of vacuum polarization effects induced by the presence of cosmic string in dS spacetime is given by the function $G_{c-dS}(x,x')$ defined by
\begin{equation}
G_{c-dS}(x,x')=\sum_{k=1}^{q-1}G_{k}(x,x') \ .
\label{Gsubk}
\end{equation}
We can see that this function is finite at the coincidence limit.

\subsection{AdS case}
Substitute \eqref{sol1-AdS} into \eqref{Green}, in Ref.~\refcite{Mello2} we have found that the expression for the Wightman functions reads:
\begin{eqnarray}
G(x,x^{\prime})&=&\frac{2q}{a^2}\left(\frac{zz^{\prime}}{4\pi}\right)^{3/2}\sideset{}{'}{\sum}_{n=0}^{\infty }\cos(nq\Delta\phi)\int_0^\infty\frac{ds}{s^4}\ e^{-({\mathcal{V}}^{2}/4s^{2})}I_{nq}\left( rr^{\prime}/{(2s^{2})}\right)\nonumber\\
&&\times I_{\nu}\left(zz^{\prime }/{(2s^{2})}\right) \ ,  \label{WF-AdS}
\end{eqnarray}
where
\begin{equation}
{\mathcal{V}}^{2}=r^{2}+r^{\prime }{}^{2}+z^{2}+z^{\prime}{}^{2}-\Delta t^{2}\ .  \label{Vcal}
\end{equation}

Once more, adopting for $q$ an integer number we can use \eqref{Sumn1} and the integral on variable $s$ can be done. For this case the Wightman function can be written as:
\begin{align}
	G(x',x)=\sum_{k=0}^{q-1}G_k(x',x) \ , \label{Greenk-AdS}
\end{align}
with
\begin{eqnarray}
G_k(x',x)=-\frac{1}{4\pi^2a^2}\frac{Q^1_{\nu-1/2}(u_k)}{(u_k^2-1)^{(1/2)}} \ , \nonumber
\end{eqnarray}
where $Q_\mu^\nu$ represents the associated Legendre function of second kind, and
\begin{eqnarray}
	u_k&=&\frac{r^2+r'^2-2rr'\cos(\Delta\phi-2\pi k/q)+z^2+z'^2-\Delta t^2}{2z z'} \ .  \nonumber\\
\end{eqnarray}

Here also we have the $k=0$ component of (\ref{Greenk-AdS}) is divergent at the coincidence limit, $x'\to x$ ($u_0\to{1}$), and represents the Wightman function in a pure AdS background.

The analysis of vacuum polarization effects induced by the presence of cosmic string in AdS spacetime is given by the function $G_{c-AdS}(x,x')$ defined by
\begin{equation}
G_{c-AdS}(x,x')=\sum_{k=1}^{q-1}G_{k}(x,x') \ .
\label{Gsubk}
\end{equation}
We can see that this function is finite at the coincidence limit.

\section{Calculation of $\langle\Phi^2(x)\rangle$}

The evaluation of the VEV of the field squared can be formally given by evaluating the Wightman function at the coincidence limit:
\begin{equation}
\langle\Phi^2(x)\rangle=\lim_{x'\to x}G(x,x') \ .\label{P2}
\end{equation}
However, in this analysis the complete Wightman functions are given by (\ref{Greenk-dS}) and (\ref{Greenk-AdS}), which are given by the sum of the Whightman in pure dS and AdS spacetimes, respectively, more the contributions induced by the cosmic string. In a compact form we may write this function as:
\begin{equation}
G(x,x')=G_{dS,AdS}(x,x')+G_{c-dS,c-AdS}(x,x') \ ,\label{Green-t}
\end{equation}
where the first term corresponds to the Wightman function in a pure dS or AdS spacetime, and the second one is the contribution induced by the presence of the cosmic string in the corresponding spacetime. Consequently by \eqref{P2} and \eqref{Green-t} we may write,
\begin{equation}
\langle\Phi^2\rangle=\langle\Phi^{2}\rangle_{dS,AdS}+\langle\Phi^{2}\rangle_{c-dS,c-AdS}\ .  \label{phi2sum}
\end{equation}
Unfortunately the above expression is divergent and some renormalization procedure is needed; however only the contribution in pure dS or AdS spacetime is divergent, so the renormalization procedure is needed only for the first contribution. As to the second contribution it is finite for points outside the string, i.e., for $r\neq0$. Because the VEV of the field squared in dS and AdS spacetimes have been analyzed by many authors, here we are mainly interested in the analysis of the quantum effects induced by the presence of cosmic string.

Due to the maximal symmetry of the dS and AdS spacetimes, the VEV of the field squared, $\langle\Phi^{2}\rangle_{dS,AdS}$, does not depend on a specific point of the corresponding spacetime; consequently, as we shall see, the contribution induced by the cosmic string, $\langle\Phi^{2}\rangle_{c-dS,c-AdS}$, will become more relevant for points near the string.

According to the previous section, we shall analyze the VEV of the field squared induced by the cosmic string for $q$ being an integer number.

\subsection{dS case}
For the dS spacetime, the contribution for VEV of the field squared due to the presence of cosmic string is given as shown below:
\begin{eqnarray}\label{Phi2-dS}
\langle\Phi^2(x)\rangle_{c-dS}=\frac{\Gamma\left(\frac{3}2+\nu\right)\Gamma\left(\frac{3}2-\nu\right)}{8\alpha^2\pi^2}	\sum_{k=1}^{q-1}\frac{P_{\nu-1/2}^{-1}(u^0_k)} {[1-(u_k^0)^2]^{1/2}} 
\end{eqnarray}
with
\begin{align}
\label{uk}
	u_k^0=-1+2(r/\eta)^2\sin^2(\pi k/q) \ .
\end{align}

In figure \ref{fig1} we have plotted the string induced part in the VEV of the field squared versus the ratio $r/\eta $ (proper distance from the string measured in units of $\alpha $) for minimally coupled scalar field ($\xi =0$) with $m\alpha =1$ (left panel) and $m\alpha =2$ (right panel). The numbers near the curves correspond to the values of the parameter $q$. For the left panel the parameter $\nu $ is real and for the right one this parameter is imaginary. In the latter case the
oscillatory behavior of the VEV is seen at large distances from the string. 
\begin{figure}[tbph]
\begin{center}
\begin{tabular}{cc}
\epsfig{figure=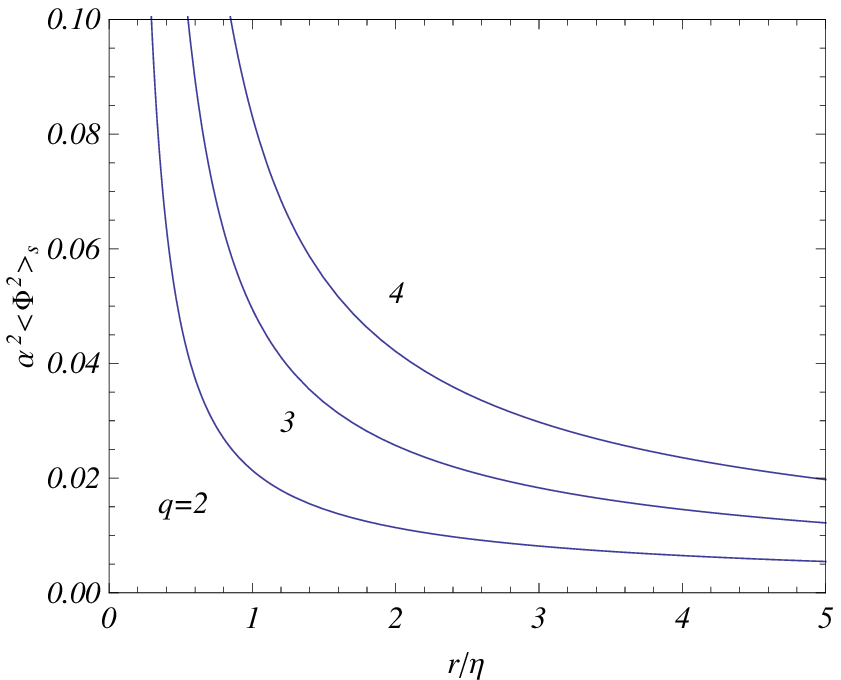,width=5.5cm,height=5.cm} & \quad 
\epsfig{figure=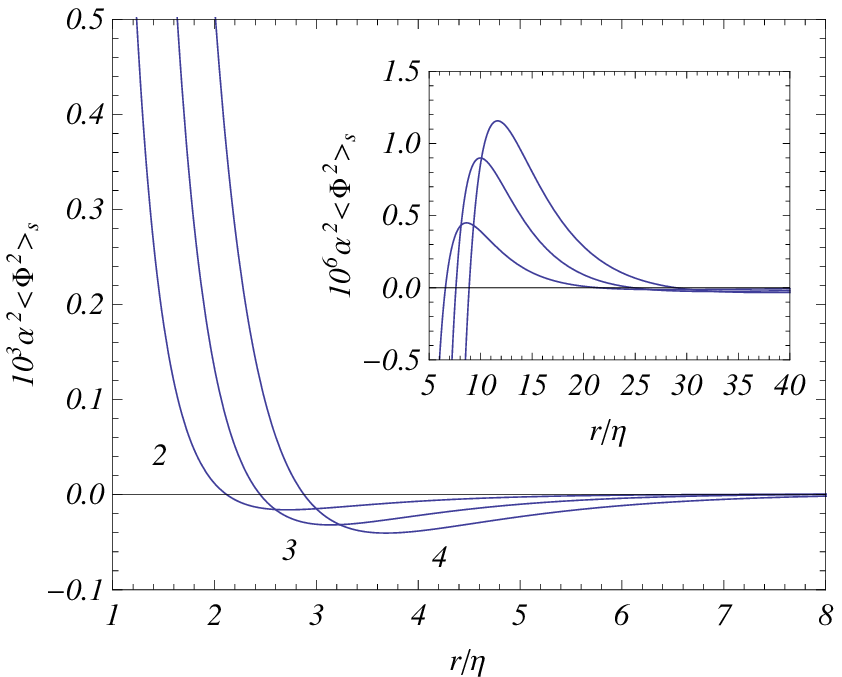,width=5.5cm,height=5cm}
\end{tabular}
\end{center}
\caption{The string induced part in the VEV of the field squared as a function of the ratio $r/\protect\eta $ for various values of the parameter $q$ for a minimally coupled scalar field with $m\protect\alpha =1$ (left panel) and $m\protect\alpha =2$ (right panel).}
\label{fig1}
\label{fig1}
\end{figure}

For the special case where the field is massless and conformally coupled with the geometry, $\xi=1/6$, one has $\nu=1/2$. Under this circumstance the Legendre function can be written in terms of elementary function as, 
\begin{align}
	P_0^{-1}(z)=\left(\frac{z-1}{z+1}\right)^{1/2} \ .
\end{align}
Consequently
\begin{equation}
\langle\Phi^2(x)\rangle_{c-dS}=\frac1{16\pi^2}\left( \frac{\eta }{\alpha r}\right)^2\sum_{k=1}^{q-1}\frac{1}{\sin^2(\pi k/q)}\ .
\end{equation}

Let us now consider the summation below
\begin{equation}
I_{N}(x)=\sum_{k=1}^{q-1}\sin ^{-N}(x+k\pi /q)\ .
\end{equation}
For even value of $N$, the above expression satisfies the recurrence relation
\begin{equation}
I_{N+2}(x)=\frac{I_{N}^{\prime \prime }(x)+N^{2}I_{N}(x)}{N(N+1)} \ .
\end{equation}
The case that we are analyzing corresponds to $N=2$. It can be shown
\begin{equation}
 I_{2}(x)=\frac{q^{2}}{\sin ^{2}(qx)}-\frac{1}{\sin ^{2}(x)}\ .
\end{equation}
Taking the limit $x\to 0$, we found $I_{2}(0)=(q^{2}-1)/3$, so
\begin{equation}
\langle\Phi^2(x)\rangle_{c-dS}=\frac{q^{2}-1}{48\pi ^{2}}\left(\frac{\eta}{\alpha r}\right)^{2}\ .  
\end{equation}
The above result is an analytical function of $q$, and by the analytical continuation it is valid for
arbitrary values of $q$.

\subsection{AdS case}
For the AdS spacetime, the contribution for VEV of the field squared due to the presence of cosmic string is given as shown below:
\begin{eqnarray}\label{Phi2-AdS}
\langle\Phi^2(x)\rangle_{c-AdS}&=&-\frac1{4\pi^2a^2}	\sum_{k=1}^{q-1}\frac{Q^1_{\nu-1/2}(u^0_k)}{((u^0_k)^2-1)^{(1/2)}} \ ,
\end{eqnarray}
with
\begin{eqnarray}
	u_k^0&=&1+2(r/z)^2\sin^2(\pi k/q) \ .
\end{eqnarray}

\begin{figure}[tbph]
\begin{center}
\epsfig{figure=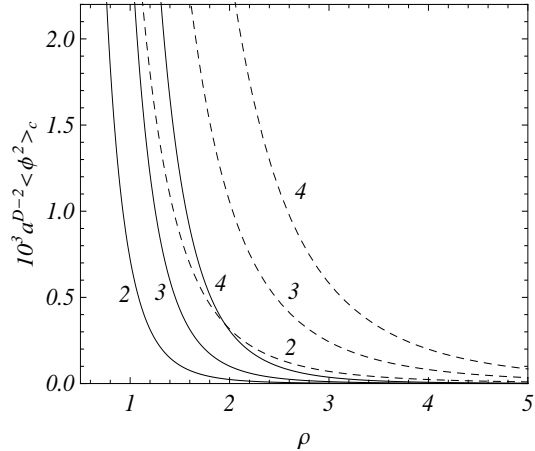,width=7.cm,height=6.cm}
\end{center}
\caption{The string-induced part in the VEV of the field squared as a function of the proper distance, $\rho=r/z$, from the string for minimally (full curves) and conformally (dashed curves) massless scalar fields. The numbers near the curves correspond to the values of the parameter $q$.}
\label{fig2}
\end{figure}

In figure \ref{fig2} we have displayed the dependence of the string-induced part in the VEV of the field squared versus the proper distance, $\rho=r/z$, from the string (measured in units of the AdS curvature radius). The graphs are plotted for minimally (full curves) and conformally (dashed curves) coupled massless scalar fields for separate values of the parameter $q$ (numbers near the curves).
\section{Conclusion}
In this paper we have analyzed the influence of a cosmic string in dS and AdS spacetimes in the evaluation of the vacuum expectation values of the quantum filed squared. The corresponding results were given by \eqref{Phi2-dS} and \eqref{Phi2-AdS}. These objectives were satisfactorily obtained because, for both spacetimes, the corresponding Wightman functions could be expressed as the sum of two terms: the first ones due to the dS or AdS backgrounds in the absence of string, and the second ones induced by the presence of the string. These facts allowed us to write the VEVs as the sum of two contributions following the same structure of the Wightman functions. Moreover, because the presence of the string does not modify the curvatures of the dS and AdS backgrounds, all the divergences presented in the calculations of VEVs of the field squared, appear only in the contributions due the purely dS or AdS space. So, the contributions induced by the string do not require renormalization. All of them are automatically finite for points outside the string. To obtain a better understanding of the quantitative behavior of both VEVs, it was provided two graphs, figures \ref{fig1} and \ref{fig2}, exhibiting their dependence on the proper distance to the string. These contributions becomes higher and higher in the regions near the string. Because the VEVs of field squared associated with pure dS and AdS spacetimes are point-independent, near the string the contributions induced by the cosmic string become more relevant than the the contributions due to pure dS or AdS spacetimes themselves; moreover, by the graphs we can also observe that the contributions induced by the string becomes higher for higher value of the parameter $q$. 

\section*{Acknowledgments}
The author thanks Conselho Nacional de Desenvolvimento Cient\'{\i}fico e Tecnol\'ogico (CNPq.) for partial financial support.

\end{document}